# Microdatabases for the Industrial Internet

K. Eric Harper, David Cox
ABB Corporate Research
940 Main Campus Drive
Raleigh, NC USA
+1 919 856 3859
{eric.e.harper, david.cox} @ us.abb.com

Thijmen de Gooijer
ABB Corporate Research
Forskargränd 7
722 26 Västerås, Sweden
+46 21 32 30 00
thijmen.de-gooijer @ se.abb.com

Johannes O. Schmitt
ABB Corporate Research
Wallstadter Straße 59
68526 Ladenburg, Germany
+49 6203 71 0
johannes.o.schmitt @ de.abb.com

## ABSTRACT
The Industrial Internet market is targeted to grow by trillions of US dollars by the year 2030, driven by adoption, deployment and integration of billions of intelligent devices and their associated data. This digital expansion faces a number of significant challenges, including reliable data management, security and privacy. Realizing the benefits from this evolution is made more difficult because a typical industrial plant includes multiple vendors and legacy technology stacks. Aggregating all the raw data to a single data center before performing analysis increases response times, raising performance concerns in traditional markets and requiring a compromise between data duplication and data access performance. Similar to the way microservices can integrate disparate information technologies without imposing monolithic cross-cutting architecture impacts, we propose microdatabases to manage the data heterogeneity of the Industrial Internet while allowing records to be captured and secured close to the industrial processes, but also be made available near the applications that can benefit from the data. A microdatabase is an abstraction of a data store that standardizes and protects the interactions between distributed data sources, providers and consumers. It integrates an information model with discoverable object types that can be browsed interactively and programmatically, and supports repository instances that evolve with their own lifecycles. The microdatabase abstraction is independent of technology choice and was designed based on solicitation and review of industry stakeholder concerns.

## Categories and Subject Descriptors
C.2.4 [**Distributed Systems**]: Distributed applications, databases; D.2.12 [**Interoperability**]: Distributed objects; D.4.2 [**Storage Management**]: Storage hierarchies; D.4.4 [**Communication Management**]: Message sending; D.4.6 [**Security and Protection**]: Access, information flow controls.

## General Terms
Management, Design, Security, Standardization.

## Keywords
Decentralized Data Management

## 1. INTRODUCTION
Low cost sensors and ubiquitous networking are enabling the next generation of industrial processing and service. This Industrial Internet shares many characteristics with mobile cloud computing, but differs in several significant ways. First, unlike on-line retail transactions, the data is not created by minimal direct human interactions with the cloud. Industrial raw measurements are created independent of hosted services making it challenging to collect and process the inputs. Second, compared to mobile phone sensor data collection, initial raw process data ownership is controlled by organizations not individuals. This increases the complexity of negotiations for who benefits from monetizing the data, especially when industrial intellectual property can be revealed by the surveillance. Finally, unlike a personal health device to log physical activities, industrial installations can have multiple vendors each with their own data representations and legacy technology stacks.

The Industrial Internet is targeted to grow by trillions of US dollars by the year 2030 [1]. The expansion brings billions of intelligent devices and connected systems. The devices have the opportunity to talk directly to one another when possible and handle much of their own computational tasks [2]. This edge computing can provide elastic resources and services, while cloud computing supports resources distributed in the core network [3].

But who benefits from this heterogeneous ecosystem? The Industrial Internet market growth will accelerate only if there is business value for both consumers and suppliers of the products and services. Devices may encounter system security challenges because they are usually deployed in the places out of rigorous surveillance and protection [4]. The suppliers must convince early adopters that their intellectual property is safe. This requires a holistic cybersecurity concept that addresses the various security and privacy risks at all abstraction levels [5].

An industrial process may be orchestrated by a single control system, but the components are selected with a best of breed strategy. Process plant design is guided in part by requirements for manufacturing precision and the cost of the individual components, bringing many different vendors into the solution space. Each component vendor has unique subject matter expertise for their equipment, making them the best analyst of the related data. Traditionally that analysis is performed only when there is a process issue, with service access to the data close to the site.

The Industrial Internet promises to increase scalability for process plant services by reducing the need to be on site. This is made possible by data collection using access from a remote location, potentially transferring the relevant measurements to the cloud. Unfortunately the dominant approach of aggregating all the data to a single datacenter significantly inflates the timeliness of analytics [6]. One approach is to establish a compromise between data duplication and the performance cost of update and select queries [7].

The necessary capabilities for Industrial Internet processing are provided in four consecutive phases: data generation, data acquisition, data storage, and data analytics [8]. A heterogeneous ecosystem comes into play in all four of these phases. For data generation, each process measurement is associated with its equipment type, converted to engineering units and validated for accuracy. Data is acquired using many different protocols and temporary storage repositories. Each component vendor has their own (legacy, hosted) platform for historical data storage and proprietary analysis applications to protect algorithms that interpret the measurements.

The keys to success for the Industrial Internet are to create value for the end users and find business models that allow various ecosystem players to co-exist and successfully co-evolve [9]. Distributed data stores are an essential component that can make this ecosystem possible.

The rest of the paper is structured as follows: Section 2 presents related work, Section 3 introduces the concept of microdatabases and defines the desired capabilities. Section 4 concludes this paper and discusses our future work.

## 2. BACKGROUND AND RELATED WORK

Technology development is more effective with small teams, where contributors focus on specific objectives independent of other activities. The risks introduced with this approach are that independent works do not integrate with each other, and the artifacts are difficult to maintain and evolve once a baseline is established. Continuous integration [10] can help, but complexity arises either in production of a monolith combining the capabilities using common libraries in a single technology stack, or with the design of interactions between different components that do not prioritize re-use.

Microservices are small, autonomous services focused on doing one thing well [11], accepting the complexity risk to enable better development productivity and faster time to market. Processing is centered on clusters of services which expose APIs (Application Programming Interface) specific to the provided capabilities, fit for purpose. A rigorously developed monolith would have the same internal library design, but without the flexibility to leverage best of breed programming languages, frameworks and deployment platforms. Furthermore, security defense-in-depth qualities may get better attention with independent microservices considering their attack surfaces and vulnerabilities.

One of the proposed characteristics of microservice architectures is decentralized data management. The idea is to divide a complex domain up into multiple bounded contexts and map out the relationships between them, enabling decentralized data storage decisions [12]. Eschewing the benefits of a single centralized repository for storage behind common services, the guidance is to choose the right persistence option for the task at hand [13]. On the other hand, related microservices can leverage the same infrastructure yet expose different information models specific to the provided services.

Decentralized data management has security qualities that enable distributed control and governance of raw data and processing results. Personal data vaults provide a privacy architecture in which individuals retain ownership of their data [14]. This concept can be extended to organizations, where configured policies for these data vaults determine which authenticated clients have access to the records, and the terms of that sharing.

What are the important capabilities and characteristics for these data vaults? These requirements are determined by exploring the scenarios and contexts [15] for data vault use, focusing on the prioritized stakeholder concerns and forces. With those inputs the system context and components are described with viewpoints and perspectives [16], highlighting the architecture qualities that address the stated concerns.

One dimension of the Industrial Internet is hosted data storage and processing coupled with analytics, known as Big Data. The NIST Big Data Public Working Group has specified an architecture framework for big data ecosystems [17]. In the framework the Data Provider exposes a collection of interfaces (or services) for discovering and accessing the data. The identified activities are collecting, persisting and retrieving data, managing PII (Personally Identifiable Information), managing metadata, configuring and enforcing access rights and supporting data discovery.

Another optional aspect of the Industrial Internet is on premise data collection and storage; for local applications, caching when communication is disrupted and eventual transfer to the cloud. The long lifecycles of traditional process industries usually result in mature operational technology on site, implemented with data historians. These MES (Manufacturing Execution Systems) serve as middleware between operational and enterprise information systems [18], largely addressing the same concerns as identified by NIST.

Common across these two ecosystems is the concept of a column store. Raw Industrial Internet data is organized as collections of key-value pairs, where the key is predominately a timestamp. These column stores are more I/O efficient for read-only queries [19] which fits well with Industrial Internet scenarios as the records are written once, not modified and read many times. An abstraction exposing a set of column stores can form the basis for our distributed data stores.

## 3. MICRODATABASE CONCEPT

The Industrial Internet can be organized in tiers, with each tier able to operate autonomously based on the available data and services. One example set of tiers is shown in Figure 1. The scope of data is limited in the lower tiers. On the other hand, local services have shortened latencies when interacting with industrial processes. There can be multiple global tiers, one for each vendor, and regional tiers are required due to country-specific regulations for data sharing cloud-to-cloud. Local tiers occur naturally from legacy operational technology deployments and device tiers arise as embedded computers expand their storage capacity and processing power.

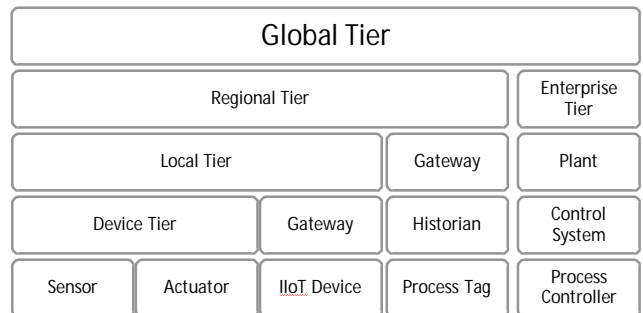

**Figure 1. Industrial Internet Tiers**

Data storage is necessary in every tier, even if only for temporary caching during communication outages. A microdatabase is defined by a standard set of APIs to securely and reliably store, manage and retrieve key-value pairs within a tier. The APIs are realized with appropriate technology available in the tier.

The repository configuration is deployed using an Industrial Internet app store (similar to mobile computing), where microdatabases are presented as peers to applications. The app store content is replicated in each tier and enables third party participation in the common ecosystem. App store transactions in a disconnected tier are replicated to other tiers when communication is re-established.

Each repository deployment can have a different information model allowing for diversity in data representation and relationships. This parallels the trend in microservices where every service has a unique set of programming interfaces and applications must know how to use them. In a similar way the microdatabase information model enables discovery and classification (tagging) of types, properties and instances.

The microdatabase system context is shown in Figure 2. The underlying data store can use any legacy or new technology as long as the API operations are supported. The microdatabase context delineates a security domain to control access and restrict operations to authorized clients. Clients must authenticate using security best practices, perhaps facilitated by the Industrial Internet app store to provide federated identity. Data exchange clients access the key-value pairs using CRUD (Create, Read, Update, and Delete). Values can be simple or structured (object) types. Some implementations may restrict updates and deletes to support data consistency goals.

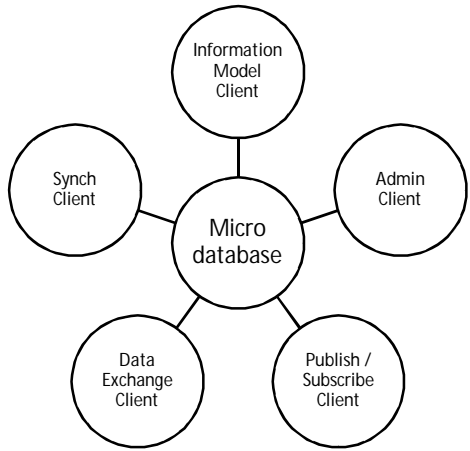

**Figure 2. Microdatabase System Context**

Similarly configured microdatabases are deployed in adjacent tiers and can be bi-directionally synchronized according filter criteria defined by the owner. The replicated repository takes on the same policies regardless of which tier the replica resides in. Each microdatabase serves as a publish and subscribe hub in its tier. Any operation on the microdatabase generates a corresponding notification published to all subscribed clients for that event.

The microdatabase owner controls the repository contents using administrative operations. The properties and configuration are declaratively specified. Column stores can be created and deleted. The repository contents can be encrypted with the owner's certificate. Each column store in a microdatabase can be connected to an ingest data source, subscribing and automatically creating records as new readings are published by the data source or by polling on a periodic basis.

Clients are provisioned and assigned to roles associated with the different interfaces, column stores, ranges of data and policies for access. Programmatic callbacks are registered for fine grained filtering of ingested, exchanged and synchronized values.

Our vision is that microdatabases can be deployed in any tier, realized with the appropriate technology choices, with synchronization provided as the only communication between tiers. Data replicated into a repository looks like ingest and triggers the associated notification events.

The following items summarize the microdatabase capabilities and their motivations.

C1. *App store deployment of configuration*. Microdatabase information model and policy definitions are deployed independent of services as first class participants in the Industrial Internet. This provides a separation of concerns between data and service ownership, and enables declarative integration of applications, services and data stores.

C2. *Integrated information model*. Asset types and instances are crucial aspects of the ecosystem: discoverable, navigable and organized independent of naming conventions. Classification of types apply to related instances and property values. Multiple information models can be federated within a tier to provide a broad view of the available storage.

C3. *Flexible classification of types, properties and instances*. Every microdatabase can invent its own type system, imposing the constraint on clients to configure and program accordingly. No different than the complexity introduced by microservice APIs, it is unrealistic that all Industrial Internet applications will agree on a common taxonomy and attributes.

C4. *Encrypted data at rest and in transfer*. Microdatabases can store encrypted data, i.e. not available to anyone outside of the provisioned users. Encryption is used both for data exchange and synchronization transactions to prevent unauthorized access.

C5. *Role-based access control configured for authenticated users*. A microdatabase imposes a security domain to protect and manage access to data. Repository owners define (select) the EULA (End User License Agreement) policies by which sharing is allowed, protecting intellectual property and sensitive information. Synchronized replicas in adjacent tiers are guarded by the same controls.

C6. *Data ingest configuration for each column store*. The Industrial Internet life blood is streams of data, including historical records that can replayed as streams. Microdatabases are populated by creating key-value pairs and data source ingest is a ubiquitous scenario.

C7. *CRUD data exchange with cascading side effects based on role*. Writing and reading key-value pairs in microdatabase column stores is the fundamental application programming model for persistence and analysis. These fine-grained transactions within a tier can be extended with programmatic callbacks configured by the repository owner, providing maximum control over the content.

C8. *Publish and subscribe notification of CRUD transactions*. Microdatabases enable clusters of processing activity in a tier by generating events associated with repository access. The events are not intended to directly share content. Instead, applications use the notifications to drive key-value pair access, similar to the MVC (Model-View-Controller) pattern.

C9. *Filtered synchronization between tiers*. Industrial Internet data is created at the network edge, yet delivers the best business value when aggregated in the cloud. Communications between the edge and cloud may not be reliable or intentionally air-gapped for security protection. Selected column store synchronization uses the network bandwidth for transferring data in bulk, and reduces the cybersecurity attack surface of a tier.

## 4. CONCLUSIONS AND FUTURE WORK

A microdatabase is a key component of a Industrial Internet ecosystem, owned and managed by a business stakeholder to provide secure storage and sharing of data within an ecosystem tier. Five sets of operations characterize microdatabase interactions with the ecosystem.

First, an integrated information model forms the basis for all interactions with the microdatabase, including design, orchestration, execution and administration. Second, key-value pairs are created, read, updated and deleted in column stores with possible configured side effects that can modify or enhance the value contents. Data source ingest is performed using create operations and application access is performed using read operations. Third, applications within an ecosystem tier subscribe to notification events published when microdatabase transactions occur, triggering actions to retrieve and process the affected content. Fourth, microdatabase contents are securely synchronized in bulk between connected tiers, using the network bandwidth to its best advantage to consolidate related content in centralized storage without losing ownership. Finally, authenticated users are authorized by the owner to configure and manage the microdatabase properties using a separate set of operations. Microdatabases enable flexible configurations of applications and data storage in a digital ecosystem, especially to integrate third parties.

The top three benefits of the microdatabase approach are to 1) reduce API (but not information model) complexity, 2) enhance privacy and security, and 3) manage connectivity. The microdatabase capabilities provide a common facility for data persistence and application notification in the Industrial Internet ecosystem, standardizing how data is managed and distributed. Applications connect to microdatabases only within a single tier, reducing the cybersecurity attack surface. Microdatabases empower and delegate responsibility to owners to protect, manage and monetize their intellectual property. Finally, given potentially unreliable communications between tiers, data vault integration adapts to and maximizes opportunities for sharing when synchronization channels are open between tiers.

As ecosystem tiers are increasingly and continuously connected, the synchronization-only communication constraint between tiers could be relaxed to allow CRUD access to microdatabases located in federated namespaces across the tiers. Microdatabases could be used to distribute content between tiers, for example, application configurations, algorithm specifications, or even the manifests and content for synchronized app stores. Finally, writing work requests to microdatabases in the local tier could synchronize with other tiers, triggering activities in adjacent tiers whose results are collected and synchronized back to the requesting tier.

## 5. ACKNOWLEDGMENTS